# Local structures and local exchange interactions in doped magnetic materials of different dimensionality

**Albert Furrer**[1,2]

[1] Laboratory for Neutron Scattering, Paul Scherrer Institut, CH-5232 Villigen PSI, Switzerland
[2] SwissNeutronics AG, Brühlstrasse 28, CH-5313 Klingnau, Switzerland

E-mail: albert.furrer@psi.ch

**Abstract**. Defects intentionally introduced into magnetic materials often have a profound effect on the physical properties. Specifically tailored neutron spectroscopic experiments can provide detailed information on both the local exchange interactions and the local distances between the magnetic atoms around the defects. This is demonstrated for manganese dimer excitations observed for the magnetically diluted three-, two- and one-dimensional compounds $KMn_xZn_{1-x}F_3$, $K_2Mn_xZn_{1-x}F_4$ and $CsMn_xMg_{1-x}Br_3$, respectively, with x=0.10. The resulting local exchange interactions deviate up to 10% from the average, and the local Mn-Mn distances are found to vary stepwise with increasing internal chemical pressure due to the Mn/Zn or Mn/Mg substitution. Our analysis qualitatively supports the theoretically predicted decay of atomic displacements according to $1/r^2$, $1/r$ and constant (for three-, two- and one-dimensional compounds, respectively) where r denotes the distance of the displaced atoms from the defect.

## 1. Introduction

Atoms in perfect crystals are arranged in a periodic lattice. However, perfect crystals do not exist in nature. All crystals have some inherent defects, *e.g.*, vacancies or substitutional atoms due to the limited purity of the material. Moreover, defects are often intentionally introduced into the crystal in order to manipulate its physical properties. Doping has become an important tool for magnetic materials in the context of high-temperature superconductivity [1] and quantum magnetism [2] as well as for semiconductors to induce ferromagnetism [3]. Defects produce local lattice distortions, *i.e.,* the atoms around the defect are displaced from their crystallographic equilibrium positions, and the exchange interactions around the defects deviate from the average exchange. Experimental information on local exchange interactions and local structures is important to understand the physical mechanisms of the defect crystals. Different experimental techniques exist to unravel local structures, the most prominent methods being x-ray absorption fine structure, nuclear magnetic resonance, nanobeam electron diffraction with a transmission electron microscope, and atomic pair-distribution function analysis, which all yield bulk information; in addition, surface-sensitive imaging methods such as scanning tunneling microscopy and atomic force microscopy are often used. All these methods provide a spatial resolution of typically 0.1 Å, and their performance can hardly be improved. It is therefore desirable to search for alternative techniques pushing the spatial resolution beyond the present limits.

Nature is rich in relationships which connect particular physical properties of materials to their structural details. One such relationship exists for the exchange coupling J of magnetic atoms, which for most materials depends on the interatomic distance R through a linear law dJ/dR as long as dR<<R. Modern spectroscopies measure exchange couplings with a precision of $dJ/J \approx 0.01$; thus, spatial resolutions of $dR \approx 0.01$ Å can be achieved. This was demonstrated for three-, two- and one-dimensional materials doped with manganese ions [4-6]. More specifically, the singlet-triplet transitions associated with manganese pairs observed by inelastic neutron scattering (INS) are found to exhibit remarkable fine structures as exemplified for the three-dimensional compound $KMn_{0.10}Zn_{0.90}F_3$ in figure 1. The fine-structure lines (indexed by m=0,1,2,...) provide detailed information on both local structural effects and local exchange interactions. As a consquence, the exchange interaction in doped materials is no longer uniformly distributed, but it can markedly deviate from the average exchange.

From the theoretical side, the defect problem was investigated by Krivoglaz [7] who predicted that the atomic displacements in three- and two-dimensional crystals decay asymptotically as $1/r^2$ and $1/r$, respectively (r denotes the distance of the displaced atoms from the defect). In one-dimensional crystals the atomic displacements do not decrease with increasing distance *r* at all. INS studies confirmed these predictions by applying statistical models [4-6] as summarized in the present work. In addition, we introduce a phenomenological model to account for the local lattice distortions, and we provide some examples to illustrate the importance of the local exchange interactions around the defects.

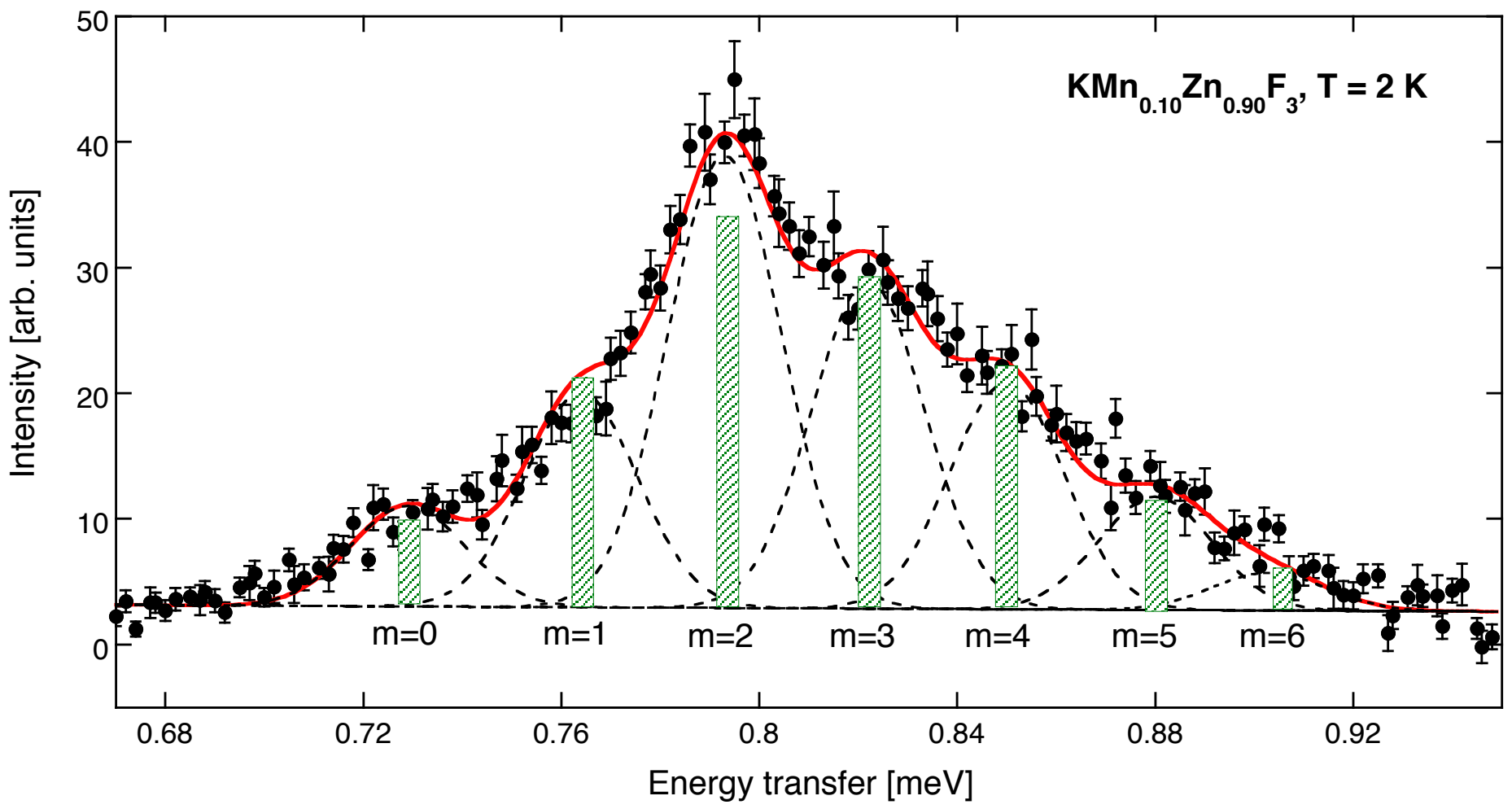

**Figure 1**. Energy distribution of the Mn singlet-triplet dimer transition observed for $KMn_xZn_{1-x}F_3$ with x=0.10 at T=2 K. The lines correspond to Gaussians resulting from a least-squares fitting procedure. The vertical bars denote the probabilities $p_m(x)$ predicted by (1) with n=26.

## 2. Theoretical background

Fig. 2 visualizes the atomic positions around isolated Mn dimers in three-, two- and one-dimensional structures relevant for the present work. For x<<1 these positions are mainly occupied by Zn or Mg atoms, which are occasionally replaced by Mn atoms according to statistics. In the following we summarize the corresponding statistical laws compatible with the theoretical predictions [7].

For the three-dimensional case as sketched in figure 2(a) for mixed Mn/Zn structures, the probability for having m Mn atoms at the n nearest-neighbor positions is given by [5]

$$p_m(x) = \binom{n}{m} x^m (1-x)^{n-m} \; . \qquad (1)$$

Because of the short-range $1/r^2$ decay law predicted by Krivoglaz [7] we do not consider further-distant-neighbor positions in (1).

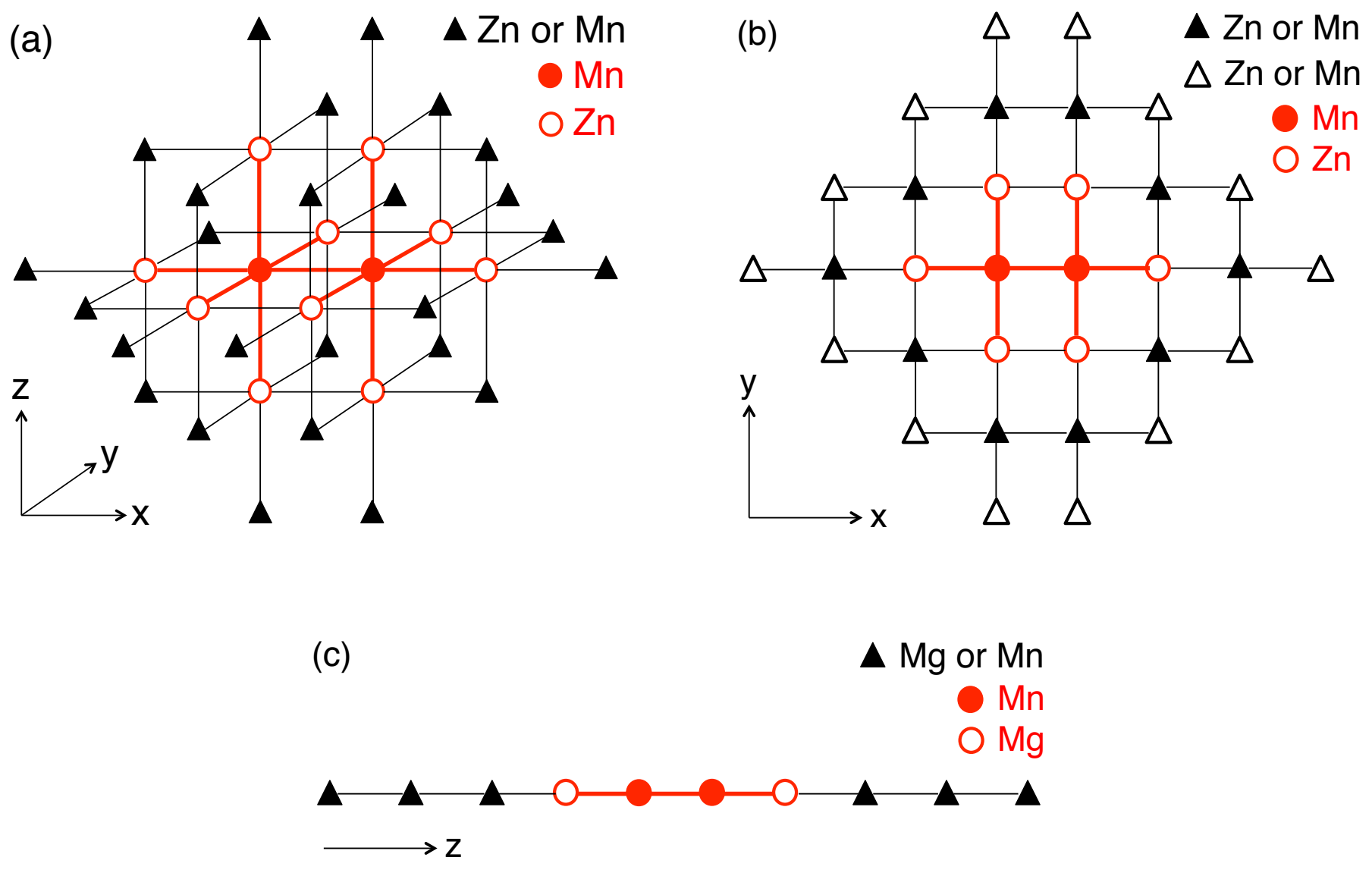

**Figure 2.** Sketch of neighboring atoms around an isolated Mn dimer marked by spheres. (a) Three-dimensional case studied for $KMn_xZn_{1-x}F_3$. The triangles denote the 26 nearest-neighbor positions. (b) Two-dimensional case studied for $K_2Mn_xZn_{1-x}F_4$. The full and open triangles denote the ten nearest-neighbor and 14 next-nearest-neighbor positions, respectively. (c) One-dimensional case studied for $CsMn_xMg_{1-x}Br_3$.

For two-dimensional systems the 1/r decay law has a long-range nature, so that the nearest-neighbor approximation is not valid. The statistical model was extended to include in (1) also the next-nearest-neighbor positions as sketched in figure 2(b), which turned out to provide a reasonable agreement with the experimental data [5].

For the one-dimensional case as sketched for a $Mn_xMg_{1-x}$ chain in figure 2(c), the probabilities $p_m(x)$ for having m Mn atoms on both sides of the central Mn dimer in a chain of length 2n are given by [4,8]

$$\begin{aligned}
p_0(x) &= (1-x)^{2n} \\
p_1(x) &= 2\binom{n}{1} x (1-x)^{2n-1} \\
p_2(x) &= \left[2\binom{n}{2} + \binom{n}{1}\binom{n}{1}\right] x^2 (1-x)^{2n-2} \\
p_3(x) &= 2\left[\binom{n}{3} + \binom{n}{2}\binom{n}{1}\right] x^3 (1-x)^{2n-3}
\end{aligned} \qquad (2)$$

etc.

The chain length 2n has to be chosen such that the sum rule $\Sigma_m p_m(x)=1$ and the condition $2n \geq m$ are satisfied. These criteria are fulfilled for n=1/x. According to Krivoglaz [7] the decay of atomic displacements is governed by the number m of Mn atoms and not by their specific arrangement in the chain.

Equations (1) and (2) were used to calculate the intensities of the fine structure in the excitation spectra of Mn dimers [4-6]. The positions of the fine-structure lines were analyzed by the spin Hamiltonian

$$H = J\mathbf{s}_1 \cdot \mathbf{s}_2 + D\left[\left(s_1^z\right)^2 + \left(s_2^z\right)^2\right] , \qquad (3)$$

where $\mathbf{s}_i$ denotes the spin operator of the Mn atoms and D the single-ion anisotropy parameter. From (3) the local exchange parameters $J_m$ can be directly determined. Through the linear law dJ/dR the local intradimer Mn-Mn distances $R_m$ can then be derived as a function of internal chemical pressure induced by the substitutional Mn atoms. Table 1 lists the parameters D and dJ/dR of the compounds considered in the present work.

**Table 1.** Single-ion anisotropy parameter D and derivative of the nearest-neighbor exchange parameter dJ/dR taken from [6]. κ denotes the local compressibility. $\Delta R_m$ corresponds to the consecutive compression of the intradimer Mn-Mn distances $R_m$ as explained in the text.

| Compound | D [μeV] | dJ/dR [meV/Å] | κ [$GPa^{-1}$] | $\Delta R_m^{obs}$ [Å] | $\Delta R_m^{cal}$ [Å] |
|---|---|---|---|---|---|
| $KMn_{0.10}Zn_{0.90}F_3$ | 5.3(2) | 3.3(6) | 0.033 | 0.0044(7) | 0.0045 |
| $K_2Mn_{0.10}Zn_{0.90}F_4$ | 5.2(2) | 2.6(4) | 0.020 | 0.0032(6) | 0.0027 |
| $CsMn_{0.10}Mg_{0.90}Br_3$ | 20.7(3) | 3.6(3) | 0.016 | 0.0022(4) | 0.0026 |

## 3. Summary of results

The data obtained for the three-dimensional compound $KMn_{0.10}Zn_{0.90}F_3$ displayed in figure 1 were subdivided into seven individual lines of Gaussian shape with equal linewidths and almost equidistant energy spacings. The probabilities $p_m(x)$ calculated from (1) with 26 nearest-neighbor positions as shown in figure 2(a) are indicated as vertical bars in figure 1. The probabilities first increase with increasing m, have a maximum for m=2, and then decrease until $p_m(x)<1\%$ for $m \geq 7$. There is a rather good agreement with the amplitudes of the individual lines with a goodness of fit $\chi^2=1.9$ [5], thus there is no need to extend the structural model beyond the nearest-neighbor positions. Actually this conclusion is in line with the theoretically predicted $1/r^2$ law for atomic displacements around a defect in three-dimensional crystals [7]. The resulting values of the local exchange parameters $J_m$ and the local Mn-Mn distances $R_m$ are displayed in figure 3(a). When going from m=0 to m=6, the consecutive internal chemical pressure exerted by the Mn substitutions results in a 0.7% compression of $R_m$ and in a 25% increase of $J_m$. The latter effect clearly shows that the exchange interaction in doped materials is no longer uniformly distributed, but it exhibits marked differences around the defects as will be further commented in section 4.

Similar results were obtained for the two-dimensional compound $K_2Mn_{0.10}Zn_{0.90}F_4$ as summarized in figure 3(b). The data analysis in terms of the probabilities $p_m(x)$ included both the 10 nearest-neighbor and the 14 next-nearest-neighbor positions as shown in figure 2(b), yielding a goodness of fit $\chi^2$=2.6 [5].

Figure 3(c) shows the results obtained for the one-dimensional compound $CsMn_{0.10}Mg_{0.90}Br_3$. The agreement between the amplitudes of the five individual lines and the calculated probabilities $p_m(x)$ defined by (1) turned out to be excellent with a goodness of fit $\chi^2$=1.2 [6], thus claiming quantitative accuracy with the theoretical prediction of a constant decay of atomic displacements for one-dimensional crystals [7].

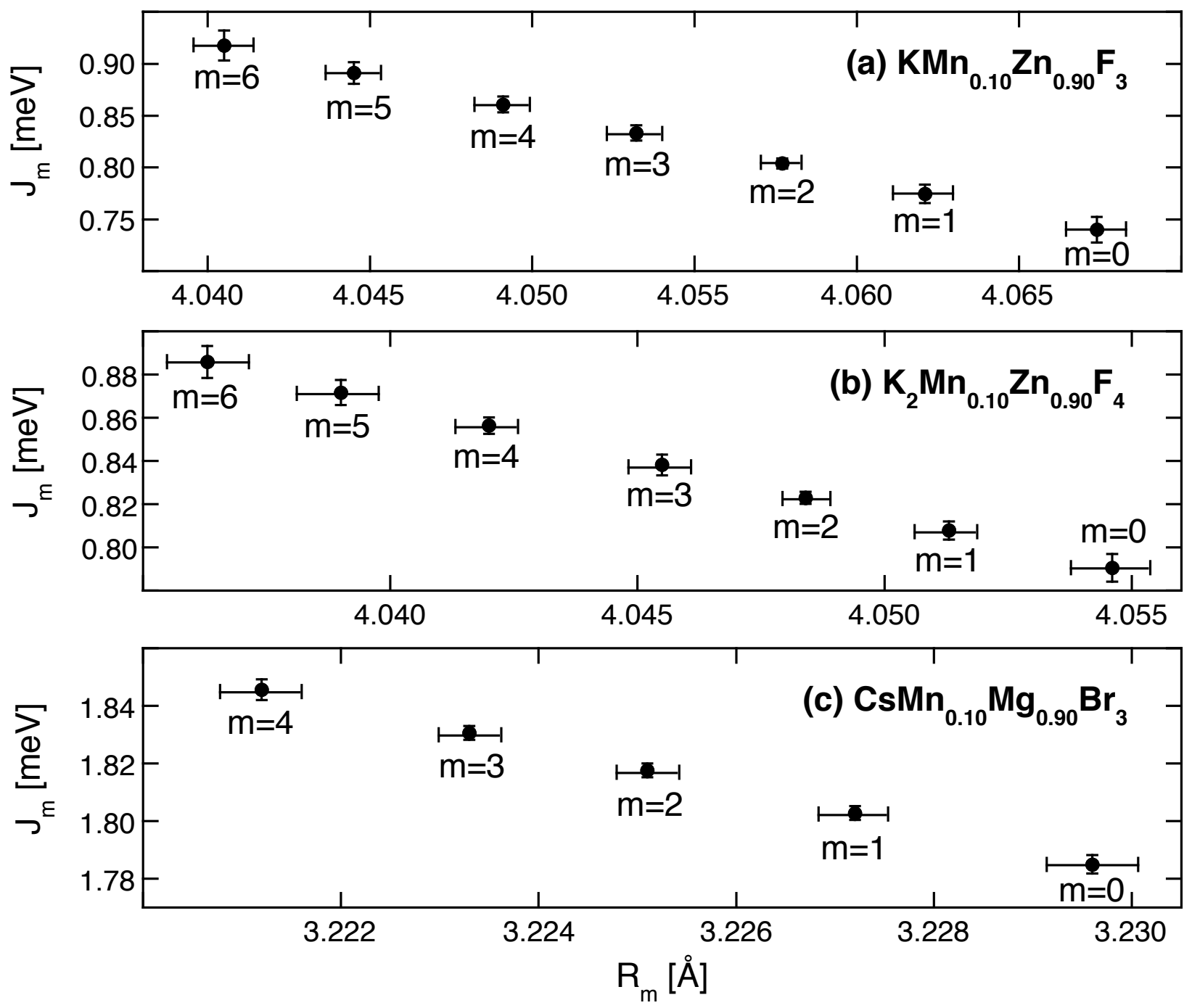

**Figure 3**. Local exchange interactions $J_m$ and local Mn-Mn distances $R_m$ derived for (a) $KMn_{0.10}Zn_{0.90}F_3$, (b) $K_2Mn_{0.10}Zn_{0.90}F_4$ and (c) $CsMn_{0.10}Mg_{0.90}Br_3$.

## 4. Concluding remarks

The analysis of the fine structure of Mn dimer excitations supports the theoretically predicted decay of atomic displacements according to $1/r^2$, $1/r$ and constant resulting from substitutional defects in three-, two- and one-dimensional crystals, respectively. The defects statistically present in the compounds clearly do not produce a continuous local dislocation pattern, but the diplacements lock in at discrete values depending on the number of defects, irrespective of their positions around the isolated Mn dimer. The substitutional atoms create a local internal chemical pressure due to the differences of the ionic radii which are $\rho_{Mn}$=0.83 Å > $\rho_{Zn}$=0.74 Å > $\rho_{Mg}$=0.72 Å for divalent ions with six-fold coordination [9]. Moreover, the nature of the internal pressure is essentially hydrostatic and most likely mediated by the elastic properties of the material. We tentatively describe the consecutive compression of the Mn-Mn distances $R_m$ displayed in figure 3 by the phenomenological law

$$\Delta R_m = R_m - R_{m+1} = \beta \cdot \Delta\rho_{ion} \cdot \kappa \ , \qquad (4)$$

where $\beta$ is a constant prefactor, $\Delta\rho_{ion}=\rho_{Mn}-\rho_{Zn,Mg}$ corresponds to the difference of the ionic radii, and $\kappa$ is the local compressibility which can be calculated from the elastic constants for the particular dimer geometries sketched in figure 2. The results are listed in table 1, which shows a good agreement between the experimental $\Delta R_m$ values taken from figure 3 and those calculated from (4) with a constant prefactor $\beta$=1.5 GPa.

Currently, investigations of quantum spin systems are of increasing interest in the search for exotic ground states. A very informative approach has been to study the modifications of the physical properties induced by the controlled introduction of defects in the materials, whose precise behavior depends on the value of the spin, the anisotropy, the dimensionality of the material, and the strengths and signs of the magnetic couplings [10,11]. The present work adds important information on the two latter quantities. We have shown that the exchange interaction in doped materials is no longer uniformly distributed, but it exhibits marked differences around the defects. The variation of the local exchange interactions around the defects manifests itself in a line broadening of the spin-wave dispersion relations as observed, *e.g.*, in INS experiments performed for $Sr_xBa_{1-x}MnO_3$ [12]. Another example concerns the spin-wave dispersion relation observed for $La_{1-x}(Ca_{1-y}Sr_y)_xMnO_3$, which was analyzed by a spin Hamiltonian with a nearest-neighbor and an unusually large fourth-nearest-neighbor exchange parameter [13]. However, the authors of [13] commented that the spin-wave data could also be reproduced by the nearest-neighbor exchange parameter $J_1$ alone, but with a random spatial distribution of different $J_1$'s resulting from the substitutional effects. The variation of local exchange interactions has also implications on the spin freezing temperature $T_g$ related to impurity induced magnetic order. As demonstrated for the compound $BiCu_{1-x}Zn_xPO_4$, the empiric law $T_g=Jx\varphi/(1+x\varphi)$ works well for small impurity contents x ($\varphi$ denotes the volume of the correlated magnetic region), but fails for x>3% due to the local variation of the exchange parameter J [11].

The spatial extension of the distorted region around the defects strongly increases with lowering the dimensionality of the spin system. In particular, these effects cannot be neglected in studies of doped spin-chain materials like $SrCuO_2$ [14] and $SrCu_2O_3$ [15] with an exchange interaction of the order of $J\approx200$ meV. Even if the magnetic couplings around the defects are modified only by a few percent, these differences match the relevant energy scale at low temperatures and therefore have to be considered in the data analysis.

**References**

[1] Bednorz J G and Müller K A 1986 *Z. Phys.* B **64** 189

[2] Sachdev S, Buragohain C and Vojta M 1999 *Science* **286** 2479

[3] Jungwirth T, Sinova J, Masek J, Kucera J and MacDonald A H 2006 *Rev. Mod. Phys.* **78** 809

[4] Furrer A, Strässle Th, Embs J P, Juranyi F, Pomjakushin V, Schneider M and Krämer K W 2011 *Phys. Rev. Lett.* **107** 115502

[5] Furrer A, Podlesnyak A, Krämer K W, Embs J P, Pomjakushin V and Strässle Th 2014 *Phys. Rev.* B **89** 144402

[6] Furrer A, Podlesnyak A, Krämer K W and Strässle Th 2014 *Phys. Rev.* B **90** 144434

[7] Krivoglaz M A 1996 *X-Ray and Neutron Diffraction in Nonideal Crystals* (Berlin: Springer)

[8] Furrer A, Juranyi F, Krämer K W, Schneider M and Strässle Th 2011 *Phys. Rev.* B **83** 024404

[9] Shannon R D 1976 *Acta Crystallogr.* A **32** 751

[10] Alloul H, Bobroff J, Gabay M and Hirschfeld P J 2009 *Rev. Mod. Phys.* **81** 45

[11] Bobroff J, Laflorencie N, Alexander L K, Mahajan A V, Koteswararao B and Mendels P 2009 *Phys. Rev. Lett.* **103** 047201

[12] Pratt D K, Lynn J W, Mais J, Chmaissem O, Brown D E, Kolesnik S and Dabrowski B 2014 *Phys. Rev.* B **90** 140401

[13] Moussa F, Hennion M, Kober-Lehouelleur P, Reznik D, Petit S, Moudden H, Ivanov A, Mukovskii Ya M, Privezentsev R and Albenque-Rullier F 2007 *Phys. Rev.* B **76** 064403

[14] Simutis G *et al* 2013 *Phys. Rev. Lett.* **111** 067204

[15] Larkin M I, Fudamoto Y, Gat I M, Kinkhabwala A, Kojima K M, Luke G M, Merrin J, Nachumi B, Uemura Y J, Azuma M, Saito T and Takano M 2000 *Phys. Rev. Lett.* **85** 1982